\documentclass[conference,letterpaper]{IEEEtran}
\IEEEoverridecommandlockouts

\usepackage{graphicx}
\usepackage{amsmath}
\usepackage{url}
\usepackage{tikz}
\usetikzlibrary{arrows.meta,positioning,calc}

\title{SCAPE: Condition-Checked Story Graphs for Interactive Video World Models}
\author{
\IEEEauthorblockN{Chanyeong Lim$^{1,*}$, Sumin Lee$^{2,*}$, Eunji Lee$^{3,*}$,
Hyuk-Jae Lee, and Nam-Joon Kim$^{\dagger}$}
\IEEEauthorblockA{
\textit{Seoul National University}, Seoul, Republic of Korea\\
$^1$\textit{Department of Computer Science and Engineering}\\
$^2$\textit{Department of Civil, Urban and Environmental Engineering}\\
$^3$\textit{Department of Mechanical Engineering}\\
\{dlacksdud2,cirtuare,mitalej,knj01\}@snu.ac.kr; hjlee@capp.snu.ac.kr
}
\thanks{$^{*}$These authors contributed equally to this work.}
\thanks{$^{\dagger}$Corresponding author: Nam-Joon Kim (knj01@snu.ac.kr).}
}

\begin{document}

\bstctlcite{scape-bib-style}

\maketitle

\begin{abstract}
Interactive video world models generate explorable scenes under real-time camera and text control. Following an authored story in these worlds requires control over which scene comes next and when the transition occurs. We present SCAPE, which guides an interactive video world through an authored story using a story graph and visual condition checks. From a short story text, SCAPE generates the graph, staged cue prompts, and scene key images. Its controller allows only the current scene's outgoing edges, stages cues to bring a transition landmark into view, and changes scene only after a separate decision model checks that the player has reached the landmark in the generated frames. At a branch, the model first identifies which landmark the player is approaching. Three independent models judge blinded frames across 47 generated worlds. On 32 linear worlds, SCAPE completes grounded transitions on 79\% of 130 planned transitions, vs.\ 62\% for a timer and 48\% for a vision-language director alone, while leaving 17 planned transitions incomplete. Relative to the timer, premature transition share is lower by 31 percentage points ($p=2\times10^{-5}$). On 15 branching worlds, SCAPE takes the player's branch at its landmark in 21 of 30 sessions, vs.\ 5 for the director alone. 
\end{abstract}

\begin{IEEEkeywords}
interactive video, world models, narrative control, story graphs,
vision-language models, evaluation
\end{IEEEkeywords}

\section{Introduction}
Interactive video world models generate an explorable scene from an image and
respond to camera and text control in real
time~\cite{lingbot2026,lingbotinfinity2026}. Building an authored narrative experience with
such a model requires control over both which scene comes next and when the
transition happens.
If the player is meant to cross a yard and enter a chapel,
the scene should change to the interior when the player reaches the chapel
door, not while the chapel is still across the yard. The video generator itself
has no representation of which scenes exist, which transitions are valid, or
when a transition condition has been met.

SCAPE encodes this structure in a directed story graph and uses it to drive
the generator at runtime. Each node is a scene with a key image, and each edge
names a \emph{transition landmark}, the object in the current scene that the
player walks up to in order to move to the next scene, such as a door or a
staircase. During play, a controller allows only the current scene's outgoing
edges and feeds staged cue prompts to bring the landmark into view. It
switches to the destination scene only after a separate decision model
checks the generated frames and confirms that the player has reached the
landmark.
Where a scene branches into multiple outgoing edges, the decision
model first identifies which landmark the player is approaching before any cue
is staged. From a few sentences of story text, SCAPE generates a complete story
world (the graph, cue prompts, and scene key images) and streams interactive
video to a browser at about 8 frames per second from one GPU.

We evaluate SCAPE across 47 generated story worlds, with three independent
vision-language models judging blinded frames at every scene transition. Our
analysis focuses on \emph{premature transitions}, in which the scene changes
before the player has reached the transition landmark on screen. We also
examine branch selection when a scene has multiple outgoing edges. On 32 linear worlds (130
planned transitions across two runs), SCAPE completes grounded transitions on
79\% of planned edges, compared with 62\% for a fixed timer and 48\% for a
vision-language director that decides scene transitions from recent frames
without a separate condition check. Relative to the timer, SCAPE lowers
premature transition share by 31 percentage points.
On 15 branching worlds, SCAPE takes the branch the player walked toward, after
reaching its landmark, in 21 of 30 sessions, compared with 5 of 30 for the
director alone, which consistently defaults to the first edge in the graph.

\section{Related Work}
\textbf{Interactive world models.} Interactive world
models extend video generation to environments that respond to user
actions~\cite{genie2024,matrixgame32026}, supporting navigation and camera
control~\cite{lingbot2026,worldplay2025,sanawm2026} as well as text-driven
events~\cite{genie32025,yume152025}.
LingBot-World~2.0~\cite{lingbotinfinity2026} pairs the generator with a VLM
director that observes recent frames and proposes events that fit the current
scene. However, this form of control does not by itself ensure that scene
transitions follow an authored narrative.
SCAPE constrains transitions to an explicit story graph
and gates every scene transition with a separate visual condition check.

\textbf{Interactive narrative control.} Branching story graphs and drama
management organize narrative progression around player
actions~\cite{riedl2005branching,mateas2005facade}. The video generator underlying
SCAPE does not provide explicit 3D player coordinates. SCAPE brings graph-based
narrative control to this setting by verifying landmark arrival directly in
the generated frames.

\textbf{Evaluating goals in generated video.} Benchmarks for generated worlds
score instruction and control adherence separately from perceptual
quality~\cite{worldscore2025,worldmodelbench2025}, verify whether specified
compositions and event sequences appear in generated
video~\cite{t2vcompbench2025,storyeval2024}, and test multi-turn
reactivity~\cite{wbench2026,worldexam2026}. Following a similar frame-based
protocol, our evaluation uses independent VLM judges to inspect the frames at
each transition and check whether the edge's condition holds when the scene
changes.

\section{SCAPE}
\begin{figure}[t]
\centering
\setlength{\unitlength}{1pt}
\scriptsize
\resizebox{\columnwidth}{!}{%
\begin{picture}(262,176)
\put(0,145){\framebox(94,20){Story graph}}
\put(0,80){\framebox(94,28){\shortstack{Graph controller\\(stage / commit)}}}
\put(0,32){\framebox(94,20){Scene change}}
\put(150,32){\framebox(94,20){Video world model}}
\put(150,128){\framebox(94,20){VLM director}}
\put(150,80){\framebox(94,24){\shortstack{Visual condition\\check}}}
\put(142,0){\framebox(106,20){\shortstack{Scene key images\\Player camera input}}}
\put(47,145){\vector(0,-1){37}}
\put(47,80){\vector(0,-1){28}}
\put(69,66){\makebox(0,0){commit}}
\put(94,42){\vector(1,0){56}}
\put(122,51){\makebox(0,0){key image}}
\put(197,52){\vector(0,1){28}}
\put(215,65){\makebox(0,0){frames}}
\put(244,42){\line(1,0){8}}
\put(252,42){\line(0,1){96}}
\put(252,138){\vector(-1,0){8}}
\put(260,90){\makebox(0,0){\rotatebox{90}{frames}}}
\put(197,128){\vector(0,-1){24}}
\put(218,116){\makebox(0,0){commit}}
\put(150,138){\line(-1,0){28}}
\put(122,138){\line(0,-1){37}}
\put(122,101){\vector(-1,0){28}}
\put(121,145){\makebox(0,0){stage}}
\put(150,87){\vector(-1,0){56}}
\put(122,77){\makebox(0,0){accepted commit}}
\put(195,20){\vector(0,1){12}}
\put(94,155){\line(1,0){103}}
\put(197,155){\vector(0,-1){7}}
\put(153,170){\makebox(0,0){current scene and its edges}}
\end{picture}}
\caption{SCAPE control loop. The director proposes actions on the current
scene's outgoing edges. Stage activates visual cues; commit is passed to
the graph controller only after a separate visual condition check on the
generated frames.}
\label{fig:architecture}
\end{figure}
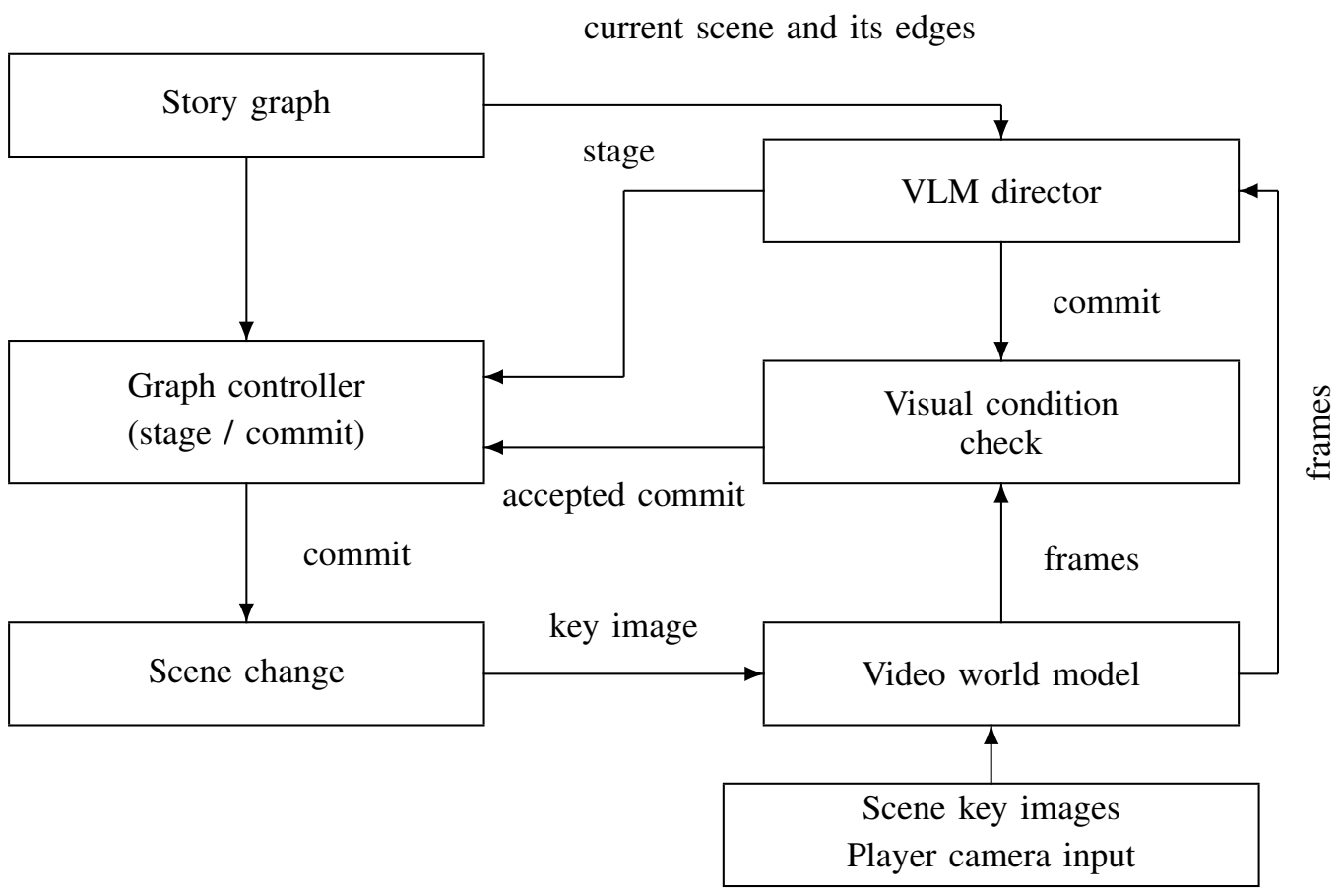

SCAPE takes a short story text as input and constructs a directed story graph
to control an interactive video world model (Fig.~\ref{fig:architecture}).
The graph specifies which scene transitions are allowed and what the player
must reach on screen to trigger each one. During play, a graph controller
tracks the current scene and stages cue prompts for outgoing edges. It
switches to the next scene only after a separate visual check confirms that
the transition condition holds in the generated frames.

\subsection{Story Graph}
\begin{figure*}[t]
\centering
\includegraphics[width=0.88\textwidth]{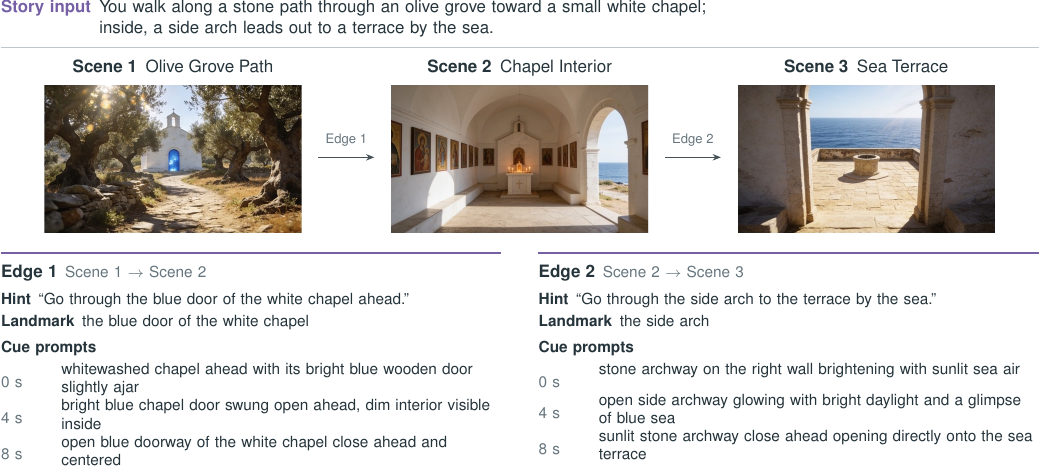}
\caption{A story graph built by SCAPE from the input story text (premise) at
the top: three scenes with their generated key images, and for each edge the
player-facing hint, the transition landmark named from it, and the three cue
prompts that are staged in the source scene.}
\label{fig:graph}
\end{figure*}

Each node in the story graph represents a scene with a generation prompt
and a key image. Each directed edge represents a transition between two scenes
(Fig.~\ref{fig:graph}).
A scene can have multiple outgoing edges, allowing the story to branch.
Each edge specifies
(i) a short player-facing hint describing the transition, (ii) a
\emph{transition landmark}, the object in the source scene that the player
walks up to in order to move to the next scene (such as a door, archway, or
staircase), and (iii) cue prompts that gradually bring that landmark into
focus in the source scene.
For example, an edge with the hint ``climb the staircase to the hallway above''
uses the staircase as its transition landmark. SCAPE checks arrival at the
staircase, and the scene transition represents passage into the hallway.

Authors can write a graph by hand or generate a complete story world from a
few sentences of story text. In the automated pipeline, Muse Spark
1.3~\cite{musespark2026} writes the scene prompts, edges, player-facing hints,
and cue prompts, Seedream 4.5~\cite{seedream2025} generates each scene's key
image, and Qwen3-VL~\cite{qwen3vl2025} extracts each edge's transition
landmark from its hint and scene description.

\subsection{Graph-Constrained Control}
The generator produces video in 16-frame chunks (1.9\,s). Between chunks, the
controller tracks the current node and supports two actions on its outgoing
edges. \emph{Stage} starts an edge's cue schedule within the current scene
(applying the three cues at 0, 4, and 8\,s and then evicting the oldest cached
frames so the final cue takes effect), and \emph{Commit} switches to the
destination scene. A VLM director (Qwen3-VL-235B) observes up to three recent frames
spaced at least 2\,s apart along with the current scene and its edges, and
proposes to stage, commit, or wait. It is instructed to request a commit once
the view described by the edge's last cue is clearly visible.

\subsection{Edge Condition Check}
Since a VLM director that reads the cue prompts often proposes a commit as
soon as the landmark appears in the distance, SCAPE gates every commit with a
separate condition check. The check receives the first frame of the scene,
the newest frame, and the transition landmark, but no cue prompts or destination
description. It places the player on a five-step scale:
\emph{not visible}, \emph{far} (less than about a quarter of the frame
height), \emph{approaching} (a quarter to a half, with several steps of ground
left), \emph{arrived} (half or more, at most a step or two left), and
\emph{passed}.
A fast decision model (Perplexity
Decider~V1.1~\cite{decider2026}) answers this scale as a single
multiple-choice question and returns option probabilities in about 0.9\,s.
After an initial 5\,s grace period, commit proposals based on frames no more
than 20\,s old are accepted only after two consecutive checks return
\emph{arrived} or \emph{passed}. Until then, the scene continues and the check
repeats on new frames.

\subsection{Choosing an Edge at a Branch}
In a scene with several outgoing edges, staging one edge's cues steers the
generated video toward that edge's landmark, so the branch must be chosen
before any cue is staged. When the director proposes to stage at a branch, the
decision model is given the transition landmarks of all outgoing edges and
asked which landmark the player is heading toward, or whether the heading is
still unclear. The controller stages the named edge and stages nothing while
the heading is unclear. Committing that edge then requires the condition check
described in Sec.~III-C.

\subsection{Scene Transition}
When a transition occurs, the controller clears the generator's context cache,
resets the camera pose, and restarts generation from the destination scene's
key image. This transition replaces continuous spatial movement through the
landmark, ensuring that play immediately lands in the destination environment;
the controller's task is to trigger the transition only when the player has
arrived at the correct landmark.

\subsection{Generator}
\begin{table}[t]
\caption{Generator configuration.}
\label{tab:runtime}
\centering\footnotesize
\setlength{\tabcolsep}{3pt}
\begin{tabular}{ll}
\hline
Backbone & LingBot-World 2.0 14B (causal)~\cite{lingbotinfinity2026} \\
Frame size & $832\times480$ \\
Chunk & 4 latent frames (16 frames), 1.9\,s on one H100 \\
Denoising & 3 steps + clean-latent cache update \\
Precision / decoder & FP8 linear layers / lightweight decoder \\
Context cache & 18 latent frames (first 6 anchored) \\
\hline
\end{tabular}
\end{table}

A session streams about 8.4 frames per second to the browser using the generator
configuration summarized in Table~\ref{tab:runtime}. Mouse movement turns the camera
horizontally around the vertical axis, using the camera tilt estimated from the
key image~\cite{geocalib2024}, while per-chunk latent normalization prevents
drift relative to the six anchored initial frames. Every chunk sent to the
client logs the active scene, the latest control action and its source, and
any chunk boundary where a transition took effect. The evaluation uses only
these client-side records and the received frames.

\section{Evaluation}
Since every transition resets generation from the destination scene's key
image, the transition always lands in the destination scene.
The evaluation
therefore asks whether the player has actually reached the transition landmark
on screen when a scene transition occurs, and whether the controller takes the
branch the player walked toward when a scene has multiple outgoing edges.

\subsection{Worlds and Input}
We wrote 48 short story texts (premises) across interior, nature, urban, and
fantasy settings: 32 \emph{linear worlds} (three scenes in sequence across two
transition landmarks) and 16 \emph{branching worlds} (a start scene with left
and right landmarks leading to different scenes). From each text, SCAPE built
one story world (graph, key images, and cues) without editing or regeneration.
The 32 linear graphs contain 65 edges, with one generated graph containing
four scenes.
One branching world was excluded before any session because a landmark was
missing from its key image, leaving 47 worlds.
Every session runs for 240\,s with scripted input. On linear worlds, the
forward key is held throughout. On branching worlds, the camera first turns
toward the left or right landmark using an angle measured from the key image,
and the forward key is then held.
Each linear world is played twice per controller and each branching
world once per branch direction.

\subsection{Controllers}
All controllers run the same graphs, cues, and generator, with all controller
parameters fixed on development recordings before generating the evaluation
worlds. A shared fallback applies only to scenes with one outgoing edge.
If cue staging has not begun, the fallback starts it after 20\,s without input
or 45\,s in the scene, without changing scenes.
We compare the following controller configurations, which differ in how they
select and commit scene transitions:
\begin{itemize}
\item \textbf{SCAPE}: the director proposes, the branch selection check
(Sec.~III-D) chooses which branch to stage, and the condition check
(Sec.~III-C) gates every commit.
\item \textbf{Timer}: stages the edge at 20\,s and requests a commit at 40\,s
without reading frames (taking effect at 46\,s after polling and chunk
alignment), controlling for elapsed time.
\item \textbf{Director only}: the director commits on its own reading of
recent frames against the edge's last cue, with no transition landmark and no
separate check, using the same model as SCAPE (Qwen3-VL-235B) following the
design of~\cite[Sec.~4.2]{lingbotinfinity2026}.
\item \textbf{Director with landmarks}: the director receives each edge's transition landmark, the five-step arrival scale, and reference frames, but decides when to transition without a separate check. The scene's first and newest frames replace its three-frame history to match the check's visual input. At a branch, it is also asked which landmark the player is approaching. Its cues and action options remain unchanged.
\item \textbf{SCAPE without check}: retains SCAPE's branch selection
(Sec.~III-D) but commits on the director's proposal without the condition
check (Sec.~III-C).
\end{itemize}
The linear worlds are played under SCAPE, Timer, Director only, and Director
with landmarks. We also test two variants of the check model, relaxing the decision
model's threshold from top choice to $P(\text{arrived}\cup\text{passed})\geq 0.35$
to test whether a looser gate recovers missed transitions, and replacing the
decision model with a generative VLM, Gemini~3.8
Flash~\cite{gemini38flash2026}, given the five-step scale as text.

The branching worlds are played under SCAPE, Director only, Director with
landmarks, and SCAPE without check. Timer is not evaluated at branches because
it does not select between outgoing edges.

\subsection{Judging and Analysis}
Three external VLM judges not used by any controller
(Gemini~3.1~Pro~\cite{gemini31pro2026}, GPT-5.6~Sol~\cite{gpt562026}, and
Claude Sonnet~5.5~\cite{claudesonnet552026}) rate blinded samples on the
five-step scale of Sec.~III-C given the scene's first frame, the frame before
transition, and the transition landmark. By majority vote, a completed
transition is labeled \emph{grounded} if the verdict is \emph{arrived} or
\emph{passed}, and \emph{premature} if it occurred before reaching the
landmark, while edges with no transition are judged on the session's final
frame. On branching worlds, the last frame before the first transition is
judged for both landmarks to check whether the controller took the branch on the
player's side after the player reached its landmark.

The unit of analysis is the story world, averaged over its two runs or two
turn inputs. We report paired differences between controllers with 95\%
bootstrap confidence intervals and exact two-sided sign-flip tests. Two primary
comparisons were specified prior to evaluating the worlds: SCAPE vs.\ Timer on
premature transition share on linear worlds, and SCAPE vs.\ Director only on
grounded branch choices on branching worlds; other comparisons are secondary.
Director with landmarks was evaluated in a single-run ablation and is compared
against the first run of the other controllers. Sessions that failed to start
or dropped before 235\,s were rerun.

\section{Results}
We first examine scene transitions and branch selection, followed by analyses
of timer settings, check models, and judge agreement.

\subsection{Condition at the Scene Transition}
\begin{table}[t]
\caption{The 65 edges of the linear worlds under each controller over two runs
(130 planned transitions per controller, majority of three judges). Scene time
shows median (quartiles) at transition.}
\label{tab:main}
\centering\footnotesize
\setlength{\tabcolsep}{3.2pt}
\begin{tabular}{lccc}
\hline
 & Director only & Timer & SCAPE \\
\hline
Transition completed & 130 & 130 & 113 \\
\quad grounded (at landmark) & 62 & 80 & \textbf{103} \\
\quad premature (before landmark) & 66 & 50 & \textbf{10} \\
\quad no majority & 2 & 0 & 0 \\
No transition within 240\,s & 0 & 0 & 17 \\
\hline
Grounded share (planned edges) & 48\% & 62\% & \textbf{79\%} \\
Scene time at transition (s) & 27 (24--31) & 46 (44--47) & 43 (34--55) \\
\hline
\end{tabular}
\end{table}

\begin{figure}[t]
\centering
\includegraphics[width=\columnwidth]{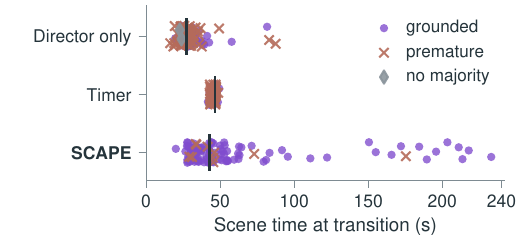}
\caption{Scene time at every transition over both runs, by controller and
verdict. Timer changes scene at a fixed time, Director only changes scene
early, and SCAPE's timing varies across scenes. SCAPE leaves 17 planned transitions incomplete within 240\,s.}
\label{fig:times}
\end{figure}

\begin{figure}[t]
\centering
\includegraphics[width=\columnwidth]{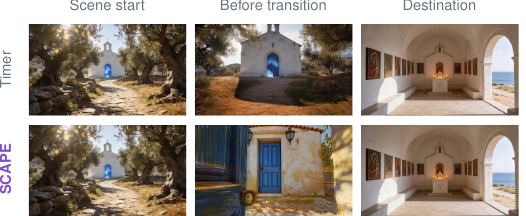}
\caption{Edge 1 of the graph in Fig.~\ref{fig:graph}. Top: Timer changes scene
with the chapel still across the yard. Bottom: SCAPE changes scene at the door.}
\label{fig:cuts}
\end{figure}

\begin{figure}[t]
\centering
\includegraphics[width=\columnwidth]{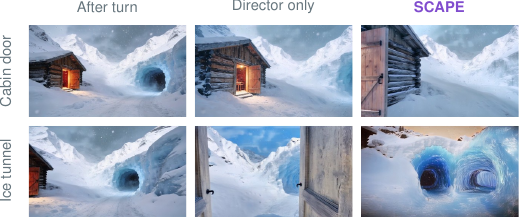}
\caption{One branching world, turning toward the cabin door (top) or the ice
tunnel (bottom). Both controllers change scene at the door for the top input.
For the bottom input, Director only again takes the cabin edge, the first in the
graph, while SCAPE changes scene at the tunnel.}
\label{fig:fork}
\end{figure}
SCAPE reduces premature scene transitions on the linear worlds
(Table~\ref{tab:main}). Under Director only, half of
all scene changes happen before the player reaches the transition landmark (66
of 130). Timer has fewer premature transitions (50 of 130) because waiting
46\,s gives the player more time to approach the landmark. SCAPE transitions
prematurely on 10 of 130 planned transitions, lowering the premature share by
$0.31$ relative to Timer (95\% CI $[0.20, 0.41]$,
$p=2\times10^{-5}$). SCAPE also raises the grounded share of planned edges to
79\% (91\% of its 113 completed transitions), compared with 62\% for Timer and
48\% for Director only (Figs.~\ref{fig:times} and~\ref{fig:cuts}).

Of the 17 edges where SCAPE did not transition within 240\,s, 11 never reached
the landmark, 2 never entered the source scene, and 4 reached the landmark near
the 240\,s cutoff before two checks completed. Of SCAPE's 10 premature
transitions, six followed video drift or collapse, and four involved a nearby
non-target doorway.

In run 1, Director with landmarks still changes scene prematurely in 28 of 65 edges,
close to Director only (32) and well above SCAPE (3). In 91 of its
92 commit calls across both world sets, the director claimed in its rationale
that the landmark already met the \emph{arrived} criteria, often citing the
active cue prompt or forward key input as evidence.

\subsection{Choosing an Edge at a Branch}
\label{sec:fork}
\begin{table}[t]
\caption{Sessions on the 15 branching worlds (a left and a right input each,
majority of three judges).}
\label{tab:fork}
\centering\footnotesize
\setlength{\tabcolsep}{2pt}
\begin{tabular}{lcccc}
\hline
 & Director & Director w/ & SCAPE w/o & SCAPE \\
 & only & landmarks & check & \\
\hline
Player's branch, grounded & 5 & 14 & 15 & \textbf{21} \\
Player's branch, premature & 10 & 11 & 10 & 1 \\
Wrong branch & 14 & 4 & 4 & 3 \\
No transition within 240\,s & 1 & 1 & 1 & 5 \\
\hline
\end{tabular}
\end{table}

On the branching worlds, where the starting scene has two outgoing edges,
SCAPE takes the player's branch with the player at
its landmark in 21 of 30 sessions, compared with 5 of 30 for Director only
(Table~\ref{tab:fork}, share higher by $0.53$, 95\% CI $[0.30, 0.77]$,
$p=0.003$). In all 29 sessions where
Director only changed scene, it took the first edge listed in the graph
whichever way the player turned, and in a 30-session run with reversed edge order
it took the newly first-listed edge in 29 sessions. SCAPE took the left branch
12 times and the right branch 13 times (Fig.~\ref{fig:fork}). In SCAPE's
wrong-branch sessions, the judges placed the player at that opposite landmark.
Both SCAPE without check and Director with landmarks reduce wrong-branch
transitions but still transition prematurely on the player's branch, which the
separate check largely removes.

\subsection{Other Timer Settings}
\label{sec:timers}
\begin{table}[t]
\caption{Landmark arrival in SCAPE sessions at scene time $T$, counting
earlier grounded transitions as reached. Ranges bound edges already
transitioned prematurely before $T$.}
\label{tab:timers}
\centering\footnotesize
\setlength{\tabcolsep}{4pt}
\begin{tabular}{lcccc}
\hline
$T$ (s) & 30 & 46 & 90 & 180 \\
\hline
Reached (\%) & 51--52 & 68--73 & 78--85 & 84--93 \\
\hline
\end{tabular}
\end{table}

As a reference for other timer settings, the judges assess the frame at candidate times $T$ in the SCAPE recordings (Table~\ref{tab:timers}). These measurements describe SCAPE's trajectories, where cues were staged dynamically, rather than the performance of additional fixed timers.

\subsection{Model Behind the Check}
\label{sec:checkmodel}
\begin{table}[b]
\caption{Variants of the edge condition check on the same 65 edges (first
run), and the median time of one check.}
\label{tab:ablation}
\centering\footnotesize
\setlength{\tabcolsep}{3pt}
\begin{tabular}{lcccc}
\hline
Check answered by & Grounded & Premature & No transition & Time \\
\hline
Decision model, top choice & \textbf{54} & \textbf{3} & 8 & 0.9\,s \\
Decision model, $P\geq0.35$ & 49 & 9 & 7 & 0.9\,s \\
Gemini 3.8 Flash & 49 & 6 & 10 & 5.2\,s \\
\hline
\end{tabular}
\end{table}
The choice of check model affects both transition outcomes and response time
(Table~\ref{tab:ablation}). On 167 development
clips labeled by the three-judge majority on the five-step scale, Gemini~3.8
Flash and Perplexity Decider~V1.1 matched 94\% and 89\% of the static-frame
verdicts. Despite its higher static-frame agreement, Gemini~3.8 Flash yields fewer grounded and more premature transitions in live play than the decision model, with slower checks and a median scene-change time of 66\,s vs.\ 43\,s. Relaxing the threshold to $P\geq 0.35$ barely reduces uncompleted
transitions while tripling premature transitions;
consistent with our analysis in Sec.~V-A, most missed transitions occur when
the generator fails to reach the landmark rather than from an overly strict
threshold.

\subsection{Judges}
Each judge alone gives the same controller ranking on both world sets.
Pairwise judge agreement is $\kappa = 0.59\text{--}0.74$ on linear samples and
$\kappa = 0.72\text{--}0.76$ on branching samples. When asked instead on run 1 whether the player
accomplished the edge's free-form hint, the judges classify 39, 29, and 11
transitions as premature for Director only, Timer, and SCAPE.

\section{Limitations}
Our evaluation measures adherence to authored visual conditions using a limited
set of repeatable, scripted inputs: forward walking and an initial turn. It
does not include human playtests or subjective ratings. Both the
runtime check and the evaluation judges use vision-language models to interpret
the same five-step scale, although the three judges come from different model
families than the decision model.

In Director with landmarks, the effect of the added landmark information
remains coupled with the switch from the director's three-frame history to the
check's first-and-newest frame pair. Cue prompts advance on a fixed schedule
rather than in response to player progress, and we did not evaluate SCAPE
without cue staging, leaving its contribution unmeasured in isolation.

SCAPE also depends on the generator to support navigation toward the landmark.
The generator sometimes fails to reach it or degrades over long sessions. Our
evaluation uses a single video backbone (LingBot-World 2.0 14B), with 0.9\,s
checks matched to its 1.9\,s chunk duration; faster generators would require
tighter latency.

\section{Conclusion}
SCAPE guides an interactive video world through an authored story by pairing
graph-restricted cue staging with a separate visual check on every scene
transition.
In a blinded evaluation on 32 linear worlds over two runs,
SCAPE completed grounded transitions on 79\% of planned edges (91\% of
completed transitions). The grounded share of planned edges was 62\% for a
timer that changed scene on a fixed schedule and 48\% for a vision-language
director that decided when to transition without a separate visual check.
On 15 branching worlds, SCAPE took the branch the player walked toward, after
reaching its landmark, in 21 of 30 sessions, whereas the director took the first
edge whenever it changed scene. The director given transition landmarks and
arrival criteria made fewer wrong-branch transitions (4 vs.\ 14), while
premature transitions remained frequent without a separate visual check.
These results show how story graphs and visual condition checks can support
authored storytelling in interactive video worlds. Reliable navigation toward
authored landmarks remains a challenge for the underlying video world model.

\section*{Acknowledgment}

This work was supported by the Institute of Information \& Communications Technology Planning \& Evaluation (IITP) grant funded by the Ministry of Science and ICT (MSIT) (IITP-0000-AI·SW Maestro).
This work was supported by IITP grants funded by MSIT (IITP-2023-RS-2023-00256081 and RS-2026-25549926) and the ANCHOR Program funded by MOE and the Seoul Metropolitan Government (2026-ANCHOR-01-110), Republic of Korea.

\bibliographystyle{IEEEtran}
\bibliography{references}

\end{document}